\documentclass[useAMS,usenatbib]{mn2e}
\usepackage{graphicx}
\usepackage{aas_macros}
\newcommand{\Msun}{M_{\odot}}
\title[Be/X-ray binaries in the Small Magellanic Cloud]{Spectral distribution of Be/X-ray binaries in the Small Magellanic Cloud \thanks{Based on astronomical observations at the European Southern Observatory La Silla in programmes 077.D--0415 and 079.D--0371.}}
\author[V. A. McBride et al.]{V. A. McBride$^{1}$\thanks{E-mail:
vanessa@soton.ac.uk (VAM)}, M. J. Coe$^{1}$, I. Negueruela$^{2}$, M.P.E.~Schurch$^{1}$ and K.E.~McGowan$^{1}$ \\
$^{1}$University of Southampton, Highfield, SO17 1BJ, United Kingdom\\
$^{2}$Departamento de F\'{\i}sica, Ingenier\'{\i}a de Sistemas y Teor\'{\i}a de la Se\~{n}al, Universidad de Alicante, Apdo. 99, 03080 Alicante, Spain }
\begin{document}

\date{Accepted 2008 April 29.  Received 2008 April 29; in original form 2008 January 22}

\pagerange{\pageref{firstpage}--\pageref{lastpage}} \pubyear{2008}

\maketitle

\label{firstpage}

\begin{abstract}
The spectral distributions of Be/X-ray binaries in the Large Magellanic Cloud and Galaxy have been shown to differ significantly from the distribution of isolated Be stars in the Galaxy.  Population synthesis models can explain this difference in spectral distributions through substantial angular momentum loss from the binary system.  In this work we explore the spectral distribution of Be/X-ray binaries in the Small Magellanic Cloud (SMC) using high signal-to-noise spectroscopy of a sample of 37 optical counterparts to known X-ray pulsars.  Our results show that the spectral distribution of Be/X-ray binaries in the SMC is consistent with that of the Galaxy, despite the lower metallicity environment of the SMC.  This may indicate that, although the metallicity of the SMC is conducive to the formation of a large  number of HMXBs, the spectral distribution of these systems is likely to be most strongly influenced by angular momentum losses during binary evolution, which are not particularly dependent on the local metallicity.
\end{abstract}


\section{Introduction}
Be/X-ray binaries are stellar systems in which a neutron star accretes material from a massive, early-type star with a circumstellar disc.  Typically they have wide, eccentric orbits with X-ray outbursts occurring when the neutron star interacts with the circumstellar material of the Be star.  Two predominant types of outburst behaviour are observed:  Type I outbursts of luminosities in the range $10^{36}$--$10^{37}$\,erg\,s$^{-1}$ last a few days and generally occur close to periastron, while Type II outbursts last much longer, reach higher luminosities ($\ga 10^{37}$\,erg\,s$^{-1}$) and show no correlation with orbital phase. 

The Small Magellanic Cloud (SMC) is home to an unusually large population of high mass X-ray binaries (HMXBs).  Based on the relative masses of the Milky Way and the SMC, there are a factor 50 more in the SMC than one would expect.  (Although the sample of HMXBs in the Milky Way is by no means complete, as can be evidenced by the recent discoveries of numerous obscured X-ray binaries \citep{WalterZuritaHerasBassani2006}, we do not expect the population to increase by a factor of 50).  Population synthesis simulations by \citet{Dray2006} show that the reduced metallicity of the SMC ($\sim$one-fifth solar) can lead to an increase in the HMXB population by a factor of three.  The metallicity is particularly relevant in influencing the evolution of massive stars, due to the impact it has on stellar winds.  The line-driven stellar winds in massive stars become weakened in low metallicity environments \citep{KudritzkiPauldrachPuls1989,VinkDeKoterLamers2001,MeynetMowlaviMaeder2006,MokiemDeKoterVink2007}, resulting in lower mass and angular momentum losses from the binary systems in which they are present.  Such angular momentum losses may be reflected in both the final mass of the compact object and the evolutionary path of the binary, i.e. whether or not the binary is disrupted by the supernova kick.  However, on its own, the reduced metallicity of the SMC cannot explain the large number of HMXB systems in this galaxy \citep{Dray2006}.  A recent increase in the star formation rate, such as required by \citet{HarrisZaritsky2004}, possibly due to tidal interaction between the SMC and its nearest neighbour the Large Magellanic Cloud (LMC), is necessary to explain this excess of HMXBs.  This is further substantiated by the results of \citet{GrimmGilfanovSunyaev2003}, which have shown that the number of HMXB systems in a galaxy can be related to the galactic star formation rate. Using indicators such as the far infrared, H$\alpha$ and ultraviolet to predict the number of HMXBs in the SMC \citep{ShtykovskiyGilfanov2005}, one finds that star formation rate alone cannot account for the number of HMXBs discovered so far in the SMC.  Hence, it seems likely that the increased star formation rate in the SMC combined with the reduced metallicity environment have given rise to the large observed population of HMXBs.

With the advent of arcsecond resolution X-ray telescopes the number of optically identified Be/X-ray binaries (all but one of the HMXBs in the SMC are Be/X-ray binaries) in the SMC has risen dramatically over the last few years.  As there are clear differences in the numbers of HMXBs between the Milky Way and SMC, which can be ascribed to metallicity and star formation, there may be other notable differences in the populations.  Most fundamentally, how do the metallicity and star formation rate reflect on the spectral distribution of the optical counterparts to the Be/X-ray binary population of the SMC?  Although their X-ray properties are well-studied, the spectral classifications of only five of the Be/X-ray binaries in the SMC are known \citep{CovinoNegueruelaCampana2001,CoeHaighLaycock2002,SchurchCoeMcGowan2007}. An accurate spectral distribution of SMC sources can provide constraints on the evolutionary models put forward to explain the spectral distribution of Be/X-ray binaries.  In addition, it may reflect the lower metallicity of the SMC and thus provide valuable insights into both binary evolution and star formation in a metal-poor environment.

\begin{figure}
\centering
\includegraphics[width=50mm,angle=270]{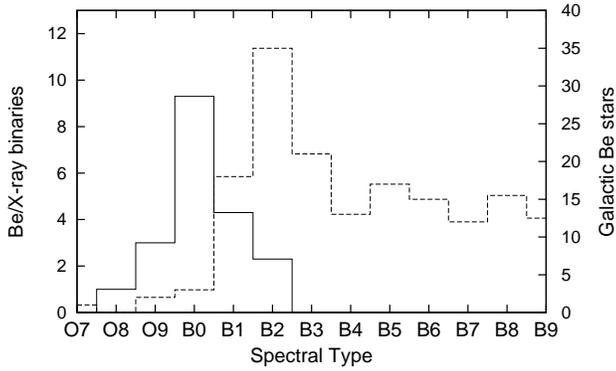}
\caption{The spectral distribution of Milky Way Be/X-ray binaries (solid) and the Milky Way isolated Be stars (dashed).}
\label{FigGal}
\end{figure}

\citet{Negueruela1998} showed that the spectral distribution of Be stars occurring in Be/X-ray binary systems is significantly different from that of isolated Be stars in the Milky Way.  Whereas isolated Galactic Be stars show a distribution beginning at the early B-types and continuing through until A0, the Be star companions of X-ray binaries show a clear cutoff near spectral type B2.  Figure~\ref{FigGal} shows the spectral distribution of Milky Way isolated Be stars (dashed histogram) compared with the Milky Way Be/X-ray binaries (solid histogram).  The Be/X-ray binary data used in the histogram are shown in Table~\ref{TabMWSpecTypes}, while the data used for the distribution of isolated Be stars are from \citet{Slettebak1982} with a magnitude limit of $V\leq6$.  Such a magnitude cutoff basically preselects the early-type Be stars and thus we expect that the true distribution may peak towards later-types than shown in Fig~\ref{FigGal}.

\begin{table}
\centering
\begin{minipage}{75mm}
\caption{The spectral types of the Milky Way Be/X-ray binaries expanded from \citet{Negueruela1998}.}
\label{TabMWSpecTypes}
\begin{tabular}{lll}
\hline
\noalign{\smallskip}
Name & Spectral Type & Ref\\
\hline
\noalign{\smallskip}
4U 0115+634 & B0.2 V & (1) \\ 
RX~J0146.9+6121 & B1 V & (2) \\
IGR~J01363+6610 & B1 & (3) \\ 
LSI+61$^\circ$~303 & B0 V & (4) \\
V0332+53 & O8--9 V & (5) \\
4U 0352+30 & B0 V & (6)\\
RX J0440.9+4431 & B0 V & (7)\\
A 0535+262 & B0 III & (8)\\
SAX J0653+0533 & B2 V -- B1 III & (9)\\
MXB 0656$-$072 & O9.7 V & (10)\\
4U 0726$-$26 & O8--9 V & (11)\\
RX J0812.4$-$3114 & B0.5 V & (12)\\
GS 0834$-$43 & B0--2 III-V & (13)\\
RX J1037.5$-$5647 & B0 V--III & (14)\\
A 1118$-$616 & O9.5 V & (15)\\
4U 1145$-$619 & B0.7 V & (16)\\
GX 304-1 & B2 V & (17)\\
2S 1417$-$624 & B1 V & (18) \\
RX J1744.7$-$2713 & B0.5 & (19)\\
Cep X-4 & B1--B2 & (20)\\
\noalign{\smallskip}
\hline
\end{tabular}
\\
(1) \citet{NegueruelaOkazaki2001};
(2) \citet{ReigFabregatCoe1997b};
(3) \citet{ReigNegueruelaPapamastorakis2005}
(4) \citet{HutchingsCrampton1981};
(5) \citet{Negueruela1998};
(6) \citet{LyubimkovRostopchinRoche1997};
(7) \citet{MotchHaberlDennerl1997};
(8) \citet{SteeleNegueruelaCoe1998};
(9) \citet{KaaretPirainoHalpern1999};
(10) \citet{pm03};
(11) \citet{NegueruelaRocheBuckley1996};
(12) \citet{MotchHaberlDennerl1997};
(13) \citet{IsraelCovinoCampana2000};
(14) \citet{MotchHaberlDennerl1997};
(15) \citet{Janot-PachecoIlovaiskyChevalier1981};
(16) \citet{FeastStoyThackeray1961};
(17) \citet{ParkesMurdinMason1980};
(18) \citet{GrindlayPetroMcClintock1984};
(19) \citet{SteeleNegueruelaClark1999};
(20) \citet{Bonnet-BidaudMouchet1998};

\end{minipage}
\end{table}

\section{Sample Selection}
The objects selected for spectroscopic follow up in this work have been selected from a sample of SMC X-ray pulsars which have positional uncertainties small enough (generally $<5^{\prime\prime}$) to allow the unambiguous identification of their optical counterparts.  The objects are drawn from \citet{CoeEdgeGalache2005}, with the exception of those that have no known well-defined X-ray position.  We also exclude SXP0.72(=SMC~X-1), which we know to be a Roche-lobe overflowing supergiant system, and SXP8.02, which is now known to be an anomalous X-ray pulsar \citep{McGarryGaenslerRansom2005}.   In addition to these objects, we include in our sample recently discovered X-ray pulsar systems with small positional errors, or previously known systems which have refined positions through detection with either \emph{Chandra} or \emph{XMM-Newton} or through previous optical follow up.  Table~\ref{TabSelect} lists the sample, along with the method used to determine the optical counterparts and any comments relating to uncertainties in the identification of the correct optical counterpart.

The resulting sample comprises 37 stars with $V$-magnitudes in the range 14--17.

\begin{table}
\centering
\label{TabSelect}
\begin{minipage}{75mm}
\caption{Sample of SMC Be/X-ray binary systems with well-defined optical counterparts.  The source name in column 1 is based on the pulse period. The column labeled {\em T} indicates the telescope used to determine the best position of the source: C -- \emph{Chandra}, X -- \emph{XMM}, A -- \emph{ASCA}, R --\emph{ROSAT}, and O -- refers to optical identification by photometry or spectroscopy of objects within the error circle.   Column 4 gives a literature reference for X-ray error circle or optical identification.}
\begin{tabular}{llll}
\hline
Source & Name & T & Ref\\
\hline
SXP0.92$^1$ & PSR~J0045$-$7319 &O & (1) \\
SXP2.37 & SMC~X-2 & O & (2) \\
SXP2.76 & RX~J0059.2$-$7138& O & (3) \\
SXP3.34 & RX~J0105.1$-$7211& A & (4)  \\
SXP6.85 & XTE~J0103$-$728 & X & (5)  \\
SXP7.78 & SMC~X-3 & C & (6) \\
SXP8.80 & RX~J0051.8$-$7231& R$^2$ & (7) \\
SXP9.13 & RX~J0049.5$-$7310& R & (4) \\
SXP15.3 & RX~J0052.1$-$7319& O & (8) \\
SXP22.1 & RX~J0117.6$-$7330& O & (9)  \\
SXP31.0 & XTE~J0111$-$7317&  O & (8) \\
SXP34.1 & RX~J0055.4$-$7210 & C,0 & (6,22) \\
SXP46.6  & 1WGA~J0053.8$-$7226& O & (10)  \\
SXP59.0  & RX~J0054.9$-$7226&  X & (11) \\
SXP65.8  & CXOU~J010712.6$-$723533& C & (12) \\
SXP74.7  & RX~J0049.0$-$7250&  O & (13)  \\
SXP82.4  & XTE~J0052$-$725 &  C & (6) \\
SXP91.1  & RX~J0051.3$-$7216&  R & (14) \\
SXP101   & RX~J0057.3$-$7325&  C & (12) \\
SXP138   & CXOU~J005323.8$-$722715& C & (6) \\
SXP140   & 2E~0054.4$-$7237 & X & (11) \\
SXP152   & CXOU~J005750.3$-$720756&C & (15, 11) \\
SXP169   & 2E~0051.1$-$7214 & R & (16) \\
SXP172   & AX~J0051.6$-$7311& X & (17) \\
SXP202   & XMMU~J005920.8$-$722316& X & (18) \\
SXP264   & AX~J0047.3$-$7312 &  X & (17) \\
SXP280   & RX~J0058.8$-$7272& X & (11) \\
SXP304   & RX~J0101.0$-$7206& C & (15) \\
SXP323   & AX~J0051$-$733 &   X & (17) \\
SXP348   & 2E~0101.5$-$7225 & C & (12) \\
SXP455   & RX~J0101.3$-$7211& X & (19) \\
SXP504   & RX~J0054.9$-$7245& C & (6) \\
SXP565   & CXOU~J005736.2$-$721934& C & (15) \\
SXP700   & CXOU~J010206.6$-$714115&C & (12) \\
SXP701   & XMMU~J005517.9$-$723853&X & (20) \\ 
SXP756   & RX~J0049.7$-$7323&  X & (17) \\
SXP1323  & RX~J0103.6$-$7201&  X & (21) \\
\hline

\end{tabular}
(1) \citet{Bell1994}; 
(2) \citet{MurdinMortonThomas1979};
(3) \citet{SouthwellCharles1996}; 
(4) \citet{CoeEdgeGalache2005}; 
(5) \citet{HaberlPietsch2008};
(6) \citet{EdgeCoeGalache2004};
(7) \citet{HaberlSasaki2000};
(8) \citet{CovinoNegueruelaCampana2001};
(9) \citet{CoeHaighLaycock2002};
(10) \citet{BuckleyCoeStevens2001};
(11) \citet{SasakiPietschHaberl2003};
(12) \citet{McGowanCoeSchurch2007};
(13) \citet{StevensCoeBuckley1999};
(14) \citet{YokogawaImanishiTsujimoto2003};
(15) \citet{MacombFoxLamb2003};
(16) \citet{CowleySchmidtkeMcGrath1997};
(17) \citet{HaberlPietsch2004};
(18) \citet{MajidLambMacomb2004};
(19) \citet{SasakiHaberlPietsch2001};
(20) \citet{HaberlPietschSchartel2004};
(21) \citet{HaberlPietsch2005};
(22) Coe 2007 priv. comm.\\
\footnotesize{1--Be ctpt to radio \& X-ray pulsar, 2--2$^{\prime\prime}$ error circle - no ambiguity in opt ctpt}
\end{minipage}
\end{table}
 
\section{Observations and Data Analysis}
 
With a view to characterising the spectral population of Be star counterparts of X-ray binaries in the SMC, we acquired spectra of 32 of the objects listed in Table~\ref{TabSelect} which do not yet have spectroscopically determined spectral types.
 
The data were acquired on the nights of 2006 September 13--15 and 2007 September 18 and 19 using the EFOSC2 faint object spectrograph mounted at the Cassegrain focus of the 3.6\,m telescope at La Silla, Chile.  We used a grism with lines ruled at 600\,mm$^{-1}$ and which covered the wavelength range $\lambda\lambda$ 3095--5085\,\AA{}.  Using a slit width of $1^{\prime\prime}$ and a CCD binning factor of $2\times 2$ throughout resulted in spectra at a resolution of 6\,\AA{}.  For fainter stars we used a lower dispersion grism ruled at 400\,mm$^{-1}$, covering the wavelength range $\lambda\lambda$3050--6100\,\AA{} and resulted in a spectral resolution of $\sim10$\,\AA{}.  Exposure times were adjusted for the stellar magnitude and observing conditions, with the average exposure time being around 1500\,s.  This was sufficient to achieve an average signal-to-noise ratio of $>100$ per pixel at $\sim4000$\,\AA{}.  Wavelength calibration was performed by taking comparison spectra of Helium and Argon lamps through the same instrumental setup.  Images of an evenly illuminated surface within the telescope dome were used for flat-fielding.
 
Data reduction, comprising bias subtraction, flat-fielding and wavelength calibration, was undertaken with version 2.12.1 of the {\em Image Reduction and Analysis Facility} (IRAF) provided by the NOAO.  
With many of our exposures being 30 to 40 minutes in duration, we applied an algorithm developed by \citet{Pych2004} to remove cosmic rays from the spectral images.  Object spectra were then inspected by eye to determine the location of the target on the CCD, after which a low order polynomial was employed to trace the spectrum across the CCD.  All spectra were extracted using the optical extraction algorithm presented by \citet{Horne1986}.  Comparison spectra were extracted along the same trace as the individual science spectra they were used to calibrate.
 
Although flux calibration was not desired for this project (spectral classification was to be performed by identifying lines rather than fitting atmospheric models), a single spectrophotometric standard star (LTT 7987) was observed to get an idea of the rough response of the telescope, grism and CCD as a function of wavelength.
 
Finally, spectra were shifted by $-150$\,km\,s$^{-1}$ \citep{Allen1973} to account roughly for the recession velocity of the SMC and hence to place spectral features at approximately the correct wavelengths.
 
\section{Classification}
 
Although many attempts have been made to apply the Morgan-Keenan (MK, \citealt{MorganKeenanKellman1943}) system of stellar classification to stars observed in the SMC \citep{Humphreys1983,AzzopardiVigneau1975}, one typically comes up against two problems.  Firstly, the low metallicity of the SMC means that the metal absorption lines are very weak, making the classification of B-type stars, which rely on metal-helium ratios, difficult.  Secondly the MK system was devised from standard stars in the Galaxy.  Classification of SMC stars straight onto the MK grid leads to contradictions between the hydrogen and metal lines and translating the MK system through a change in metallicity is a non-trivial task.

The defining characteristics of B-stars in the MK system are that they show neutral He lines in their spectra, but no ionised He (as found in O-stars \citealt{JaschekJaschek}), with the maximum strength of the He~{\footnotesize I} lines being reached around spectral class B2.  Hydrogen line strengths peak at A2, so the H and He lines show opposing trends through the B-type sequence.  The metal lines of Si~{\footnotesize II, III, IV}, Mg~{\footnotesize II} and ratios of these to He~{\footnotesize I} (and to a lesser extent the C~{\footnotesize II}, O~{\footnotesize II} and N~{\footnotesize II}) are used for spectral classification of B-type stars.  
 
In early B-types the N~{\footnotesize II}~$\lambda 3995$/He~{\footnotesize I}~$\lambda 4009$ and Si~{\footnotesize IV}~$\lambda 4089$/He~{\footnotesize I}~$\lambda\lambda4026,4121$ line ratios, which become larger towards more luminous stars, can be used to distinguish luminosity effects.  Also, the He~{\footnotesize I}~$\lambda 4121$/He~{\footnotesize I}~$\lambda 4144$ ratio, strengthening towards more luminous stars, is used for luminosity class determination.  Towards later B-types the profiles of the hydrogen lines are used, with the line profiles being narrower out of the main-sequence.  It is worth pointing out that, for Be stars, use of the Balmer lines in stellar classification is limited due to the frequent infilling of these lines by the circumstellar disc.
 
To overcome the setbacks of applying the MK system in the SMC, \citet{Lennon1997}, using high signal-to-noise spectra of SMC supergiants, devised a system for classification of stars in the SMC.  This system is ``normalised'' to the MK system such that, moving through the range of spectral types, stars in both systems exhibit the same trends in their line strengths.  This classification method was implemented by \citet{EvansHowarthIrwin2004} and \citet{EvansLennonSmartt2006} in recent spectroscopic surveys of massive stars in the SMC, the LMC and the Galaxy.  For classification of the spectra in this work, we have used the temperature criteria as set out by \citet{Lennon1997} and utilised by \citet{EvansHowarthIrwin2004} (see Table~\ref{TabClassLennon}).

\begin{table}
\centering
\caption{Classification criteria for B-type stars in the SMC, from \citet{Lennon1997,EvansHowarthIrwin2004}.}
\label{TabClassLennon}
\begin{tabular}{ll}
\hline
\noalign{\smallskip}
Criterion & Spectral type\\
\hline
\noalign{\smallskip}
He~{\footnotesize II} $\lambda4200\sim$He~{\footnotesize I} $\lambda$4143 & O9\\
He~{\footnotesize II} $\lambda\lambda4686,4541$ present, $\lambda4200$ weak  & B0\\
He~{\footnotesize II} $\lambda\lambda4200,4541$ absent, $\lambda4686$ weak & B0.5\\
He~{\footnotesize II} $\lambda4686$ absent, Si~{\footnotesize IV} $\lambda\lambda4088,4116$ present & B1\\
Si~{\footnotesize IV} $\lambda4116$ absent, Si~{\footnotesize IV} $\lambda4088<$O~{\footnotesize II} & B1.5\\
Si~{\footnotesize IV}, Si~{\footnotesize II} absent, Si~{\footnotesize III} $\lambda4553>$ Mg~{\footnotesize II} $\lambda4481$ & B2\\
Si~{\footnotesize III} $\lambda 4553\sim$ Mg~{\footnotesize II} $\lambda4481$ & B2.5\\
Si~{\footnotesize III} $\lambda 4553<$ Mg~{\footnotesize II} $\lambda4481$ & B3\\
Si~{\footnotesize III} absent, Si~{\footnotesize II} $\lambda4128/4132<$ He~{\footnotesize I} $\lambda4121$ & B5\\
He~{\footnotesize I} $\lambda4121<$ Si~{\footnotesize II}$<$ He~{\footnotesize I} $\lambda4143$ & B8\\
Mg~{\footnotesize II} $\lambda4481\leq$ He~{\footnotesize I} $\lambda4471$ & \\
Mg~{\footnotesize II} $\lambda4481>$ He~{\footnotesize I} $\lambda4471$ & B9\\
Fe~{\footnotesize II} $\lambda4233<$ Si~{\footnotesize II} $\lambda4128/4132$ & \\
\hline
\end{tabular}
\end{table}

Wherever possible, we used the trends in the Si/He and H/He line ratios to estimate the luminosity classes of the observed stars.  However, in many cases the Si lines were not evident due to the combined effects of low metallicity, extreme rotational line broadening intrinsic to Be stars and the 6\,\AA{} spectral resolution of the data.  Furthermore, it was difficult to tell whether the He lines were suffering from the same emission veiling as manifest in many of the Balmer lines.  So, for the purposes of cross-checking these luminosity class estimates, we used the apparent V-magnitudes of the Be stars together with the distance modulus of the SMC (18.9, \citealp{HarriesHilditchHowarth2003}) to determine whether the absolute magnitude of the Be star was consistent with the estimated luminosity class at the given spectral type (using absolute magnitudes of OeBe stars from \citealt{Wegner2006}).  Iterating through these procedures of firstly identifying the spectral type, secondly using spectral features to determine the luminosity class and finally validating this luminosity class against the absolute magnitude resulted in a consistent set of spectral and luminosity classifications for these stars.

\begin{figure}
\includegraphics[width=85mm]{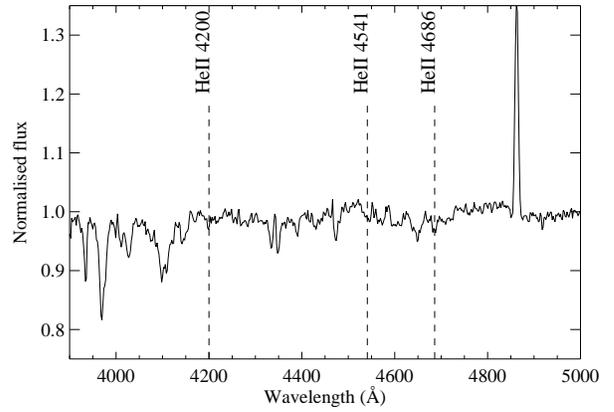}
\caption{The spectrum of SXP701, with defining features marked on the plot.}
\label{FigSxp701}
\end{figure}

This method is illustrated by the spectrum of SXP701, which appears in Fig.~\ref{FigSxp701}.  The presence of ionised He in the spectrum, at the wavelengths $\lambda 4686$\,\AA{} and $\lambda\lambda4200,4541$\,\AA{} allows us to determine the spectral type as O9.5.  The luminosity criteria at this spectral type are the line ratios of Si~{\footnotesize IV} $\lambda4089$/He~{\footnotesize I} $\lambda\lambda4026,4121,4144$ and Si~{\footnotesize II} $\lambda4116$/He~{\footnotesize I} $\lambda4121$.  As the Si lines are weakened by the lower metallicity environment of the SMC, we use the strengthening He~{\footnotesize II} $\lambda4686$ alongside the C~{\footnotesize III}/O~{\footnotesize II} blend to estimate a luminosity class of V \citep{WalbornFitzpatrick1990}.  The apparent magnitude of $V=15.87$ confirms this luminosity class, as a star of higher luminosity class would need to be of later spectral type, and this can certainly be discounted on the basis of the He~{\footnotesize II} lines. 

\begin{figure}
\includegraphics[width=80mm]{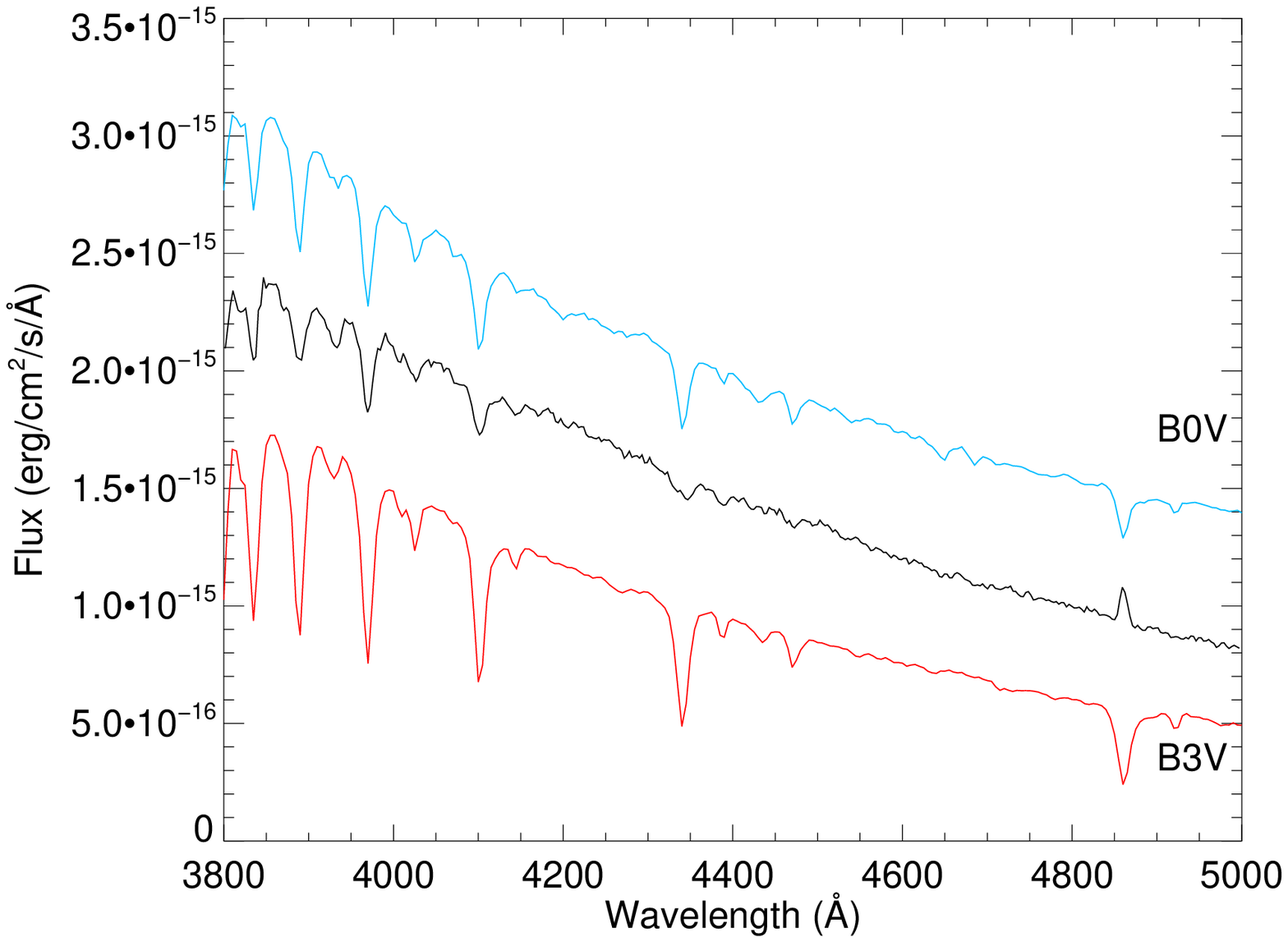}
\caption{The spectrum SXP138 plotted between the template spectrum of a  \hbox{B0~V} star and the template spectrum of a \hbox{B3~V} star.}
\label{FigSxp138}
\end{figure}

For fainter stars which were observed with the lower dispersion grism, we employed two methods that would give us an idea of the general spectral type of the star, rather than attempting to use very broad and shallow spectral lines to pinpoint the exact spectral type.  The dereddened spectra  (using an extinction of $E(B-V)=0.16$\footnote{This is the average extinction for optical counterparts of Be/X-ray systems in the SMC as measured by \citet{ZaritskyHarrisThompson2002} at http://ngala.as.arizona.edu/dennis/smcext.html}) were compared to template spectra \citep{Pickles1998} of different spectral types, using the general continuum and line features to constrain the spectral type.  In addition, the He~{\footnotesize I}~$\lambda\lambda4009,4026$/H$\epsilon$ line ratio, which decreases rapidly through the B main sequence and is hardly influenced by emission from the circumstellar disc, was used to gauge the rough spectral class.  Once again, the apparent magnitude of the star was used to confirm the spectral and luminosity class.  Figure~\ref{FigSxp138} demonstrates the value of this approach for SXP138.  It is clear just from the slope of the spectrum that the star is earlier than B3, but probably later than B0 due to the lack of He~{\footnotesize II} absorption features in the spectrum.

\section{Results}

\begin{table*}
\centering
\begin{minipage}{120mm}
\caption{Properties of SMC Be/X-ray binaries.  Source names are related their X-ray pulse periods.  In the {\em Ref} column, O refers to colours from \citet{UdalskiSoszynskiSzymanski1998}, M refers to colours from \citet{Massey2002}, while Z refers to colours from \citet{ZaritskyHarrisThompson2002}.  The {\em Grism} column refers to the ruling, in l/mm of the grism used in each case.  The {\em Spec} column lists the classifications determined from blue spectra in this work unless otherwise noted.  The luminosity class, as determined from the spectrum in the case of 600\,l/mm data, and as determined from the absolute magnitude in the case of 400\,l/mm data, of each source is given in the {\em Lum} column.}
\label{TabColours}
\begin{tabular}{lrrrrllll}
\hline
\noalign{\smallskip}
Source & $V$ & $\Delta V$ & $B-V$ & $\Delta(B-V)$ & Ref &Grism  & Spec & Lum\\
\hline
\noalign{\smallskip}
SXP0.92 & 16.18 & 0.02 & -0.21 & 0.03 & O & 600& B0.5--B2 & IV--V\\
SXP2.37 & 16.38 & 0.02 & 0.02 & 0.04 & Z & 600& O9.5 & III--V\\
SXP2.76 &  14.01 & 0.08 & 0.06 & 0.09 & Z &600&  B1--B1.5 & II--III\\
SXP3.34 &  15.63 & 0.03 & -0.01 & 0.05 & O &600&  B1--B2 & III--V\\
SXP6.85 & 14.59 & 0.02 & -0.08 & 0.02 & M & 600& O9.5--B0 & IV--V\\
SXP7.78 & 14.91 & 0.02 & 0.00 &  0.03 & Z &600&  B1--B1.5 & IV--V \\
SXP8.80 &  14.87 & 0.12 & -0.27 & 0.13 & O &600&  O9.5--B0 & IV--V\\
SXP9.13 &  16.51 & 0.02 & 0.01 & 0.04 & O & 400& B1--B3 & IV--V\\
SXP15.3 &  14.67 & 0.04 & -0.01 & 0.05 & O & 600& O9.5--B0 & III--V \\
SXP22.1 &  14.18 & 0.03 & -0.04 & 0.04 & Z & 600& O9.5--B0 & III--V\\
SXP31.0 &  15.52 & 0.03 & -0.10 & 0.04 & Z & 600& O9.5--B1$^{1}$ & V\\
SXP34.1 &  16.78 & 0.03 & -0.12 & 0.04 & Z & 400& B2--B3 & IV--V\\
SXP46.6 &  14.72 & 0.03 & -0.07 & 0.03 & Z & 600& O9.5--B1 & IV--V\\
SXP59.0 &  15.28 & 0.01 & -0.04 & 0.02 & O & 600& O9 & V\\
SXP65.8 &  15.64 & 0.03 & -0.12 & 0.03 & M &600& B1--B1.5$^{2}$ & II--III\\
SXP74.7 &  16.92 & 0.06 & 0.09 & 0.01 & O & 400& $\sim$B3 & V\\
SXP82.4 & 15.02 & 0.02 & 0.14 & 0.03 & O & 600& B1--B3 & III--V\\
SXP91.1 &  15.05 & 0.06 & -0.08 & 0.06 & Z & 600& B0.5 & III--V\\
SXP101  &  15.67 & 0.15 & -0.05 & 0.15 & M & 600& B3--B5$^{2}$ & Ib--II?\\
SXP138  & 16.19 & 0.12 & -0.09 & 0.12 & Z & 400& B1--B2 & IV--V\\
SXP140  & 15.88 & 0.03 & -0.04 & 0.03 & Z & 600& B1 & V \\
SXP152  & 15.69 & 0.03 & -0.05 & 0.12 & Z & 600& B1--B2.5 & III--V\\
SXP169  & 15.53 & 0.02 & -0.05 & 0.04 & Z & 600& B0--B1 & III--V\\
SXP172  & 14.45 & 0.02 & -0.07 & 0.02 & O & 600& O9.5--B0 & V\\
SXP202  & 14.82 & 0.02 & -0.07 & 0.01  & O & 600& B0--B1 & V\\
SXP264  & 15.85 & 0.01 & 0.00 & 0.01 & Z &600& B1--B1.5 & V \\
SXP280  &  15.64 & 0.03 & -0.12 & 0.04 & Z & 600& B0--B2 & III--V \\
SXP304  &  15.72 & 0.01 & -0.04 & 0.02 & O & 600& B0--B2 & III-V\\
SXP323  & 15.44 & 0.04 & -0.04 & 0.05 & O & 600& B0--B0.5$^3$ & V\\ 
SXP348  & 14.79 & 0.01 & -0.09 & 0.01 & O & 600& B0.5 & IV--V\\
SXP455  &  15.49 & 0.02 & -0.07 & 0.05 & O & 600& B0.5--B2 & IV--V\\ 
SXP504  &  14.99 & 0.01 & -0.02 & 0.01 & O & 600& B1 & III--V\\
SXP565  & 15.97 & 0.02 & -0.02 & 0.04 & O & 600& B0--B2 & IV--V\\
SXP700  & 14.60 & 0.02 & -0.08 & 0.02 & M & 600& B0--B0.5$^2$ & III--V\\
SXP701  & 15.87 & 0.05 & 0.15 & 0.05 & M & 600& O9.5 & V\\
SXP756  &  14.98 & 0.02 & 0.05 & 0.03 & O & 600& O9.5--B0.5 & III--V\\ 
SXP1323 &  14.65 & 0.02 & -0.11 & 0.03 & Z & 600& B0 & III--V\\
\hline
\end{tabular}
\begin{footnotesize}(1)\citet{CovinoNegueruelaCampana2001} (2)\citet{SchurchCoeMcGowan2007} (3)\citet{CoeHaighLaycock2002} \end{footnotesize}
\end{minipage}
\end{table*}

\begin{figure}
\centering
\includegraphics[width=50mm,angle=270]{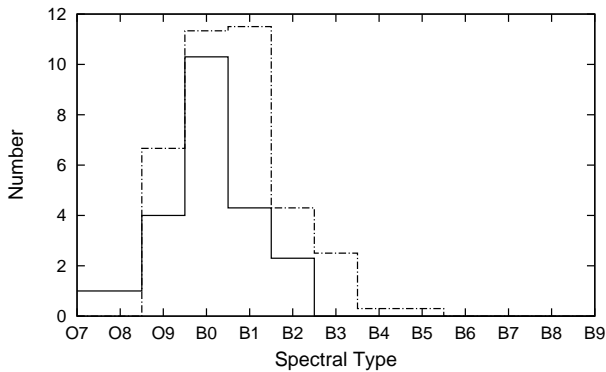}
\caption{Spectral distribution, as determined from high signal-to-noise blue spectra, of Be/X-ray binaries in the SMC (dot-dashed) as compared the distribution of Be/X-ray binaries in the Galaxy (solid).}
\label{FigSmcSpecDist}
\end{figure}

The spectral distribution of SMC Be/X-ray binaries is shown in Fig.~\ref{FigSmcSpecDist} with individual spectral types given in Table~\ref{TabColours}.  Our procedure for the binning of stars where we were unable to constrain the spectral type to one subtype involved allocating the object fractionally to the spectral classes within the estimated spectral range of the source.  For example, SXP82.4 which can be constrained to the B1--B3 range translates into the histogram as $\frac{1}{3}$ B1, $\frac{1}{3}$ B2 and $\frac{1}{3}$ B3. 

The distribution shows a similarity to the spectral distributions of the Galactic \citep{Negueruela1998} and LMC \citep{NegueruelaCoe2002} Be/X-ray binaries.  The spectral distribution of Be/X-ray binary counterparts in the SMC peaks at spectral type B1, compared to the LMC and Galaxy distributions, which peak at B0.  The Galactic and LMC distributions show a sharp cutoff at B2, whereas there are 5 SMC objects with \emph{possible} spectral types beyond B2, but as one can see from Table~\ref{TabColours} the exact spectral type cannot be determined with certainty in these cases.  A Kolmogorov-Smirnov test of the difference between the SMC and Galactic distributions gives a K-S statistic $D = 0.22$, indicating that the null hypothesis (which is that the two distributions are the same) cannot be rejected even at significances as low as 90\%.  Hence, it is likely that both Galactic and SMC Be/X-ray binary counterparts are drawn from the same population.

The most noteworthy selection effect, which may mask the real distribution of spectral types in the SMC, is magnitude related:  fainter stars, which are most likely to be those of later B-types, are the most difficult to observe and classify.  Table~\ref{TabColours} shows that the latest spectral types observed among the SMC systems are those of SXP74.7 ($\sim$B3) and SXP101 (B3--B5).  Although SXP74.7 is faint, spectral types of stars much fainter than SXP101 have been confidently determined. Thus, there seems very little evidence that magnitude related selection effects have influenced our sample.  A second factor reduces the number of optical counterparts included in our sample in the first place:  this is related to those SMC X-ray pulsars which do not as yet have an identified optical counterpart.  This is predominantly an effect of large uncertainties on the positions of pulsars as measured by non-imaging X-ray satellites, and we do not expect the absence of these objects in our sample to introduce spectral type selection effects.

\section{Discussion}
A number of authors \citep{RappaportVandenHeuvel1982,VanBeverVanbeveren1997} suggest the possibility that Be stars in X-ray binaries may be a product of binary evolution.  During mass transfer the secondary star, or at least its outer layers, may be spun up \citep{Packet1981}.  As it is well-known that the Be phenomenon is closely associated with high stellar rotational velocities \citep{PorterRivinius2003} it is not far-fetched to suppose that such binary mass transfer may rejuvenate the secondary star and, somehow, give rise to the Be phenomenon.   Studies of Be stars in clusters \citep{McSwainGies2005} indicate that most cases of the Be phenomenon may be brought about by spin-up processes in close binaries, rather than Be stars being born as fast rotators or being spun-up in the final stages of their main sequence lives.

Such a model of rejuvenation may also account for the narrow spectral distribution of Be/X-ray binaries.  In population synthesis models \citet{PortegiesZwart1995} includes the effects of both supernova kick velocities and angular momentum loss through the $L_2$ point.  He finds that this angular momentum leakage is just as influential in restricting the final number of Be binary pulsars as the effect of a supernova kick and, in addition, that angular momentum loss restricts the spectral distribution of the Be star companions.

\citet{PortegiesZwart1995} shows clearly that a higher loss of angular momentum restricts the mass of the Be star counterparts to $>8\Msun$, i.e. to spectral types earlier than ~B2 V \citep{Allen1973}.  This is caused by the fact that binary systems holding late-type stars, which have smaller mass ratios, tend to undergo a spiralling-in effect when mass is lost through the $L_2$ point \citep{PolsCoteWaters1991}.  
In these cases the components may merge or form a common envelope.  Whatever the eventuality in such a case, it is clear that a Be neutron star system will not be the outcome.

\citet{NegueruelaCoe2002} have shown that Be/X-ray binaries in the LMC follow the same distribution as those in the Milky Way, and this work (see Fig.~\ref{FigSmcSpecDist}) shows that the spectral distribution of Be/X-ray binaries in the SMC is also consistent with that of the Milky Way systems.

Although no major spectral survey of Be stars in the SMC has been performed, preliminary results \citep{MartayanBaadeHubert2006} obtained from photometric colours of $\sim7700$ emission line stars indicate that the spectral distribution of isolated Be stars in the SMC is similar to that in the Galaxy.   
Hence, at this time we cannot exclude the possibility that isolated Be stars in the SMC may have a different spectral distribution from those in the Milky Way.

\begin{figure}
\centering
\includegraphics[width=80mm]{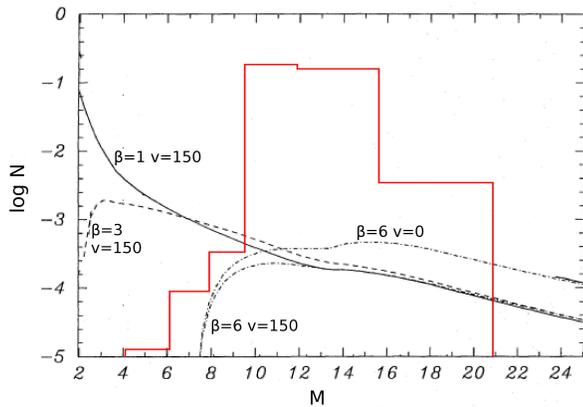}
\caption{The absolute normalised number distribution of Be stars with a neutron star companion for four evolutionary scenarios. $v$ represents the supernova kick velocities (in km\,s$^{-1}$) while $\beta$ represents the angular momentum lost from the binary system. The solid histogram represents the spectral distribution of SMC Be/X-ray binaries from this work (spectral type/mass calibration from \citet{Zombeck2007}, arbitrarily scaled along vertical axis) superimposed.  Original figure from \citet{PortegiesZwart1995}.}
\label{FigPort7}
\end{figure}

Figure~\ref{FigPort7} shows the arbitrarily scaled spectral distribution of SMC Be/X-ray binaries superimposed on the predicted spectral distribution of Be/X-ray binaries (from \citealt{PortegiesZwart1995}).  As with the Milky Way and LMC distributions, the SMC distribution cuts off around spectral type B2 ($\sim8$\,M$_{\sun}$), indicating that there may be significant angular momentum losses in the binary system prior to the Be/X-ray binary evolutionary phase.  A possible interpretation of the fact that there is no significant metallicity dependence of the spectral distributions of Be/X-ray binaries is that the angular momentum is lost through mechanisms other than the stellar winds of early-type components of these systems.  For example, as \citet{PortegiesZwart1995} infers, the mass may be lost through the $L_2$ point during the initial phase of mass transfer which serves to spin up the star to Be star status.  It is also worth noting that there are no SMC Be/X-ray binaries with masses greater than $\sim22$\,M$_{\sun}$.  This may be due to the fact that heavier systems go on to become supergiant X-ray binaries, which have lifetimes significantly shorter than the Be/X-ray systems and are consequently less abundant (there is only one in the SMC -- SMC X-1).

\section*{Acknowledgments}

VAM thanks the National Research Foundation, South Africa and British Council for financial support.  IN acknowledges support from the Spanish Ministerio de Educaci\'on y Ciencia through grant AYA2005-00095.

\bibliographystyle{mn2e}

\label{lastpage}

\end{document}